\documentclass[prl,aps,amssymb,showpacs,twocolumn]{revtex4}
\usepackage{amsmath}
\usepackage{amssymb}
\usepackage{amsthm}
\usepackage{amsfonts}
\usepackage{enumerate}
\usepackage{latexsym}
\usepackage{bm}
\usepackage{graphicx}
\usepackage{subfigure}
\newcommand{\beq}{\begin{equation}}
\newcommand{\eneq}{\end{equation}}

\input{epsf}

\begin{document}

\tolerance 10000

\newcommand{\vk}{{\bf k}}

%\draft

\title{Hole Spin Helix: Anomalous Spin Diffusion in Anisotropic Strained Hole Quantum Wells}

\author{Vincent E. Sacksteder IV$^{1}$}
\email{vincent@sacksteder.com}
\affiliation{Institute of Physics, Chinese Academy of Sciences, Beijing 100190, China}
\affiliation{Division of Physics and Applied Physics, Nanyang Technological University, 21 Nanyang Link, Singapore 637371}

\author{B. Andrei Bernevig}
 \affiliation{Princeton Center for Theoretical
Science, Princeton, NJ 08544} \affiliation{$^2$Department of
Physics, Princeton University, Princeton, NJ 08544}

\begin{abstract}
We obtain the spin-orbit interaction and spin-charge coupled transport equations of a two-dimensional heavy hole gas under the influence of strain and anisotropy.  We show that a simple two-band Hamiltonian can be used to  describe the holes.  In addition to the well-known cubic hole spin-orbit interaction, anisotropy  causes a Dresselhaus-like term, and strain causes a Rashba term.   
We discover that strain can cause  a shifting symmetry of the Fermi surfaces for spin up and down holes.  We predict  an enhanced spin lifetime associated with a spin helix standing wave  similar to the Persistent Spin Helix which exists in the two-dimensional electron gas with equal Rashba and Dresselhaus spin-orbit interactions.  These results may be useful both for spin-based experimental determination of the Luttinger parameters of the valence band Hamiltonian  and for creating long-lived spin excitations.
 \end{abstract}

\date{\today}

\pacs{72.25.Dc, 72.10.-d, 73.50.-h, 73.63.Hs}
\maketitle

Systems with spin-orbit interactions have generated great academic and practical interest  \cite{wolf2001,murakami2003,kane2005,bernevig2006,Sattler10} because they  allow for
purely electric manipulation of the electron spin \cite{nitta1997,grundler2000,kato2004}, which could be of use in areas ranging from spintronics to quantum computing. However spin-orbit interactions have also the undesired effect of causing spin decoherence \cite{dyakonov1986}. Recently a new mechanism by which a system can sustain both strong spin-orbit interactions and long spin relaxation times has been proposed \cite{bernevig2006A}.  In properly tuned systems a non-decaying spin density standing wave can be excited.  This  Persistent Spin Helix (PSH) has been observed  through spin transient experiments in electron doped GaAs quantum wells. \cite{weber2007,koralek2008}

In electron doped samples the PSH occurs when the Rashba and Dresselhaus spin-orbit interaction strengths are tuned to match each other. At equal strength the  spin dynamics conserve an $SU(2)$ triplet of spin operators, two of which describe spin standing waves, while the third describes  a uniform spin density that is selected and preserved by the tuned spin-orbit interaction.  The triplet's infinite lifetime is obtained by tuning the spin-orbit interaction  to have a constant phase independent of electron momentum, in which case the electron spin structure is independent of momentum and conserved under scattering.
 In particular,  the  Rashba and linear Dresselhaus terms are proportional to $k_- = k_x - \imath k_y$ and to $k_+ = k_x + \imath k_y$ respectively, so when they are at equal strength the total spin-orbit interaction has constant phase, producing long-lived spin excitations.

 The experimental discovery of the PSH in the 2-D electron gas raises the question of whether it exists  in other systems. Recently one of us predicted that tuned topological insulators can host PSHs with very long lifetimes \cite{Sacksteder12}.   Here we examine 2-D hole gases under strain,  calculate the  spin orbit interaction of the  heavy holes, and find Rashba and Dresselhaus-like terms caused respectively by  applied strain and  anisotropy.   When fine-tuned properly the spin orbit term has constant phase, producing long-lived spin  helices aligned with  the strain axis and an anomalous enhancement of the  spin lifetime.  
 These results apply also to other systems which like holes are four-fold degenerate at the $\Gamma$ point, such as the  metallic phase predicted in pyrochlore iridates. \cite{Yang10,Wan11,Krempa12}  PSHs are a general phenomenon that can be realized in diverse systems with a wide variety of tuning parameters.

Hole-doped quantum wells  are sensitive to applied strain and to anisotropy, which are key to the spin physics uncovered here.  Unlike the electron gas,  in the 2-D hole gas the spin components $S_x, S_y, S_z$ are decoupled at leading order if there is neither strain nor anisotropy \cite{hughes2006}.  This is due to the holes' spin-orbit interaction  $H_{\text{SO}}$, which determines the  couplings between spin components.  In the unstrained isotropic hole gas  this operator has a cubic form with f-wave symmetry $H_{\text{SO}} = \alpha (k_+^3 \sigma_- - k_-^3 \sigma_+)$ and hence vanishes when integrated over the isotropic Fermi surface. However recent experiments performed by attaching a piezo to hole-doped GaAs samples have  revealed that strain can be used to cause large changes in the spin orbit interaction \cite{Kolokolov99, Kraak04, Habib07, shabani2008}.  The experimental results largely confirm the standard Kane and Luttinger $\vec{k} \cdot \vec{p}$  Hamiltonian which predicts that strain and anisotropy  substantially  deform the two heavy hole Fermi surfaces \cite{winkler2000,Kolokolov99,Kraak04, shabani2008}. For a certain critical value of the strain field, the  surfaces meet at two special "touching points", which causes an experimentally confirmed \cite{Kraak04} magnetic breakdown of Shubnikov-de Haas (SdH) orbits.  These touching points hint at interesting spin dynamics, because similar degeneracies are seen in the Fermi surfaces of electron-doped systems when they are tuned to produce PSHs.

\begin{figure}[t]
\centering
\includegraphics[width=8.5cm, height=5.7cm]{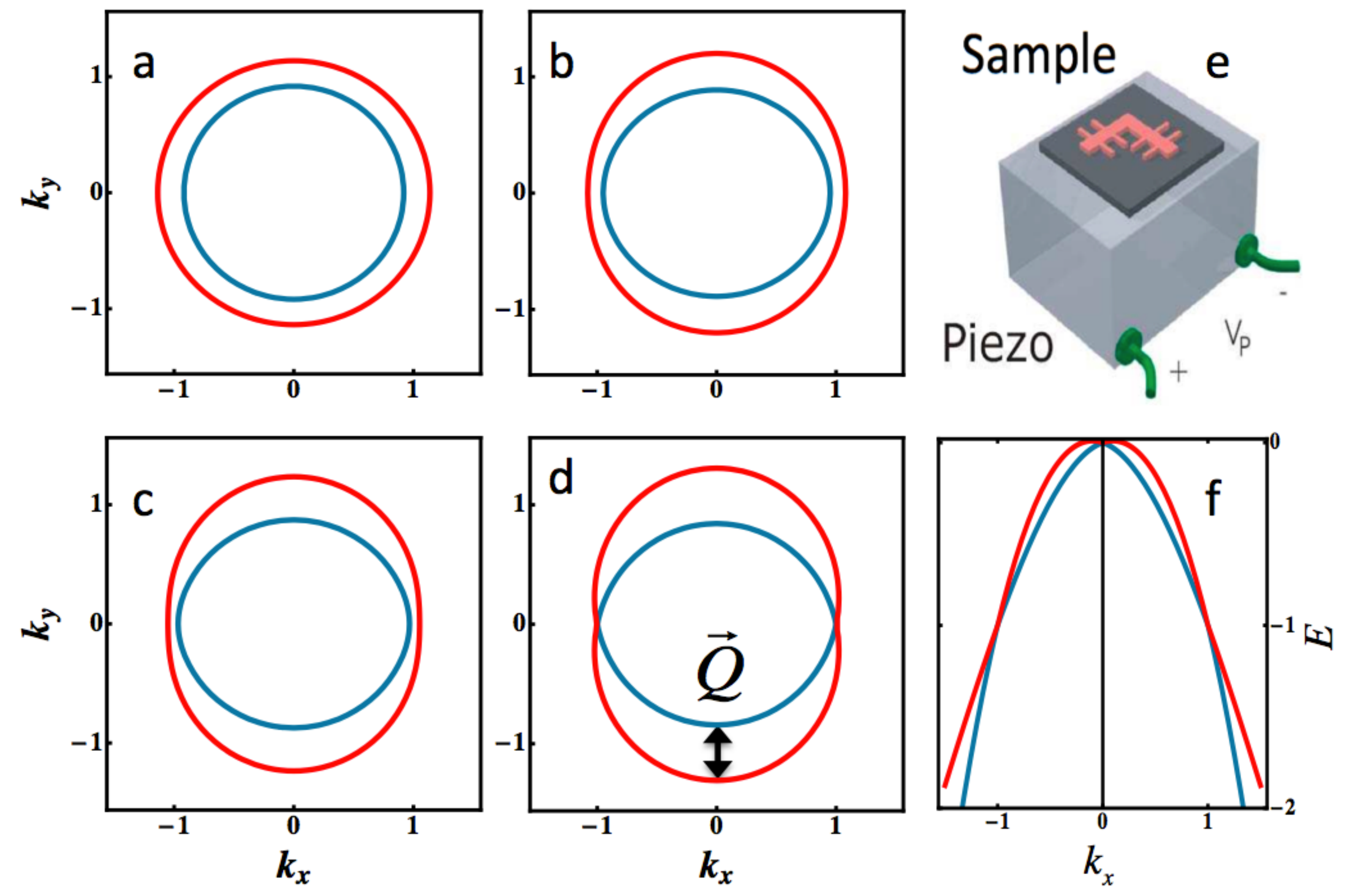}
\caption{
Tuning the strain produces touching points in the heavy hole  Fermi surfaces, which are shown at zero strain  $\beta^2=0$ (pane a) and  intermediate strains  (b) $0.4 \cdot k_F^2$ and (c) $0.6\cdot k_F^2$.  Finally pane (d) and (f) show  critical strain $\beta^2 = k_F^2$, with two touching points on the $k_x$ axis.     At critical strain shifting by $\pm \vec{Q}$ places one band on top of the other, as seen in pane (d).       The strain is oriented along the $x$ axis, and  $\gamma_2/\gamma_3 = 1$.  Identical results are obtained when the sign of $\gamma_2 / \gamma_3$ is reversed and the strain axis is rotated by 90 degrees.    $E_F = 1$ and  $\alpha = 0.2$.   Pane (e): schematic of a sample glued on a strain-generating piezo as used in Ref. \cite{Habib07}.     }\label{PSHHoleGasStrain}
\end{figure}

 Usually the heavy hole Fermi surfaces and their deformation under strain are modeled with considerable accuracy within the $4$-band Luttinger model \cite{Luttinger56} or the $8$-band Kane  Hamiltonian \cite{Kane57}.  These models unfortunately obscure  the spin-orbit interaction between the two heavy hole states and prohibit analytical calculation of  the spin-charge dynamics.  We will therefore focus only on the heavy holes, and will make explicit their spin-orbit interaction $H_{\text{SO}}$, which is simply the off-diagonal element of the two-band effective Hamiltonian which governs the heavy holes.  Various previous works have developed two-band  models of the heavy holes \cite{Kleinert07, Raichev08, Liu08, Wang12, Dollinger13}; ours distinguishes itself  by including strain.

We derive our two-band Hamiltonian from the 4-band Luttinger Hamiltonian $H_0$, which describes the total angular momentum $j = 3/2$ band that lies nearest to the Fermi surface. There are four states: two heavy holes with $j_z = \pm3/2$ and two light holes with $j_z = \pm 1/2$.   Following common practice, we choose the bulk Hamiltonian appropriate for crystal growth along the high-symmetry $z$ (001) axis, and we take the hole carrier concentration to be small enough that only the first 2-D subband in the quantum well contributes to transport.  \cite{Kamburov12,  Kolokolov99, Kraak04, Dai09, Dollinger13, Scholz13, Ekenberg85, Liu08, Lu05,winkler2000}   We include a strain field $\epsilon_{ij}$ using the Bir-Pikus strain Hamiltonian $H_{\epsilon}$ \cite{Bir74}, and we model  the quantum well with a confinement potential $V_c$ and a small charge asymmetry $V_E  = - e E z$;      
\begin{eqnarray}
&H = H_0  + H_{\epsilon} + V(z), \; V(z) = V_c + V_E 
\\
&H_0 = \frac{1}{2m}\left({\gamma_1 + \frac{5}{2} \gamma_2}\right) |\vec{k}|^2  - \frac{\gamma_2}{m} k_i^2 S_i^2  -2 \frac{\gamma_3}{m} \{k_i,k_j\} \{S_i, S_j\} \nonumber
\\
& H_\epsilon= a \epsilon_{ii} + b \epsilon_{ii} S_i^2 + d \epsilon_{ij} \{S_i, S_j\} \nonumber
\end{eqnarray}
\noindent  $\vec{S}$ is a spin $3/2$ matrix.  The double index implies summation (in the anticommutators, do not sum over ${i=j}$), and  $\{A,B\} =\frac{1}{2}(AB+BA)$.      We will show that hole spin physics  is a sensitive measure of anisotropy in the valence band, which is parameterized by three Luttinger parameters $\gamma_i$. $\gamma_1,\gamma_2$  control the hole masses  along the $z$ axis, while  $\gamma_1, \gamma_3$ control the masses along the $[111]$ axis.   Both these parameters and  the strain deformation potentials $a,b,$ and $d$   have widespread applications and are reported in standard reference works. \cite{LandoltBornstein17,Vurgaftman01}   

The Luttinger Hamiltonian has the most general form possible for a $\vec{k} \cdot \vec{p}$ model with four degenerate bands (angular momentum $j = 3/2$) in a crystal with both cubic discrete symmetry and time reversal symmetry.  Zinc blende semiconductors are not  symmetric under inversion and therefore possess only tetrahedral symmetry which is a subgroup of  cubic symmetry, but this asymmetry  is weak in the bulk \cite{Lipari70}.  Several works have examined terms beyond the Luttinger Hamiltonian and developed their effects on heavy holes \cite{winkler2000,Winkler03, Raichev08,Dollinger13}.  Here we retain only the Luttinger Hamiltonian and use an explicit term   $V_E =  - e E z$ to break inversion symmetry.

The spin-orbit physics can be illuminated by breaking the Hamiltonian explicitly into the heavy hole sector $j_z = +3/2, \, -3/2$ and the light hole sector $j_z = +1/2, \, -1/2$:
\begin{eqnarray}
H &=& V(z)+ \begin{bmatrix} 1 \cdot k_{HH} & U_{HL} \\ U_{HL}^\dagger & 1 \cdot k_{LL}   \end{bmatrix}, \;  U_{HL} = \begin{bmatrix} S & R^* \\ R & - S^*   \end{bmatrix}
\nonumber \\
S &=& -  d ( \imath \epsilon_{zy} -  \epsilon_{zx}) - \sqrt{3} \gamma_3 k_- k_z/m
\\ \nonumber 
R&=&- \frac{\sqrt{3} }{4m} ( k_+^2 (\gamma_2 + \gamma_3) + k_-^2(\gamma_2 - \gamma_3)      - 2 \gamma_3 \beta^2 e^{\imath 2 \theta} )
\end{eqnarray}
\noindent The in-plane strain is encapsulated in a  magnitude $\beta^2$ and orientation $\theta$  which are set by $ \beta^2 e^{\imath 2 \theta} = b \frac{m}{\gamma_3}(\epsilon_{xx}- \epsilon_{yy})+ \imath \frac{2m}{\sqrt{3} \gamma_3} d \epsilon_{xy} $.  The kinetic terms $k_{HH}$ and $k_{LL}$ are respectively equal to $(\gamma_1 + \gamma_2)k^2/2m + (\gamma_1 - 2 \gamma_2)k_z^2/2m $ and  $ (\gamma_1 - \gamma_2)k^2/2m + (\gamma_1 + 2 \gamma_2)k_z^2/2m $, plus a strain-induced constant splitting.  $S$ couples holes with the same sign of $j_z$, while $R$ couples  holes with opposite sign.  

This explicit representation reveals that there is no direct interaction either between  the   $j_z = \pm 3/2$ heavy holes or  between the $j_z = \pm 1/2$ light holes.   In consequence the spin-orbit interaction between the heavy holes is proportional to $R^*$.  \footnote{This result remains true at leading order in the Hamiltonian of Ref. \cite{Dollinger13}, where inversion asymmetry was added to the Luttinger Hamiltonian. }   This is an exact result.  It  informs us that when  $R=0$ there is a degenerate point in the dispersion, where the two heavy hole bands meet \cite{Kolokolov99}.  In fact, addition of a tuned strain field \emph{generically} creates  two such degenerate points on the Fermi surface.  Figure \ref{PSHHoleGasStrain} illustrates this in the particular case of compression along the $x$ axis.  In this special case the two degenerate points lie on the same axis and occur when  the strain is tuned for resonance with the Fermi momentum $\beta = k_F$.

We procede by deriving the exact two-band effective Hamiltonian which controls the heavy holes, $H_{HH} = V(z) + 1 \cdot k_{HH} + U_{HL} G_{LL} U_{HL}^\dagger$, where $G_{LL} = (E - k_{LL} - V(z))^{-1}$ is the light hole Green's function.  (See the supplementary material for an expanded derivation.) This can be rewritten as 
\begin{eqnarray}
H_{HH} &=&\left(
                  \begin{array}{cc}
                   \hat{k}_{HH} & H_{\text{SO}} \\
                     H_{\text{SO}}^* & \hat{k}_{HH} \\
                  \end{array}
                \right), \; H_{\text{SO}} = -[G_{LL}, S] R^*
                \nonumber \\
\hat{k}_{HH} &=& k_{HH} + V(z) + S G_{LL} S^* + R G_{LL} R^*
\end{eqnarray}
\noindent The commutator $[G_{LL}, S] =-\sqrt{3} \gamma_3 k_-/m [G_{LL}, k_z] $ is insensitive to strain and its phase is set by $\imath k_-$.  Therefore the  phase of the spin-orbit interaction is determined by $\imath k_- R^*$.  We make this exact result explicit by writing the spin-orbit interaction as $H_{\text{SO}} =  - \imath \alpha \frac{ 2  m }{\sqrt{3} \gamma_3} k_- R^*  = { \imath  \alpha}( k_-^3 (1 + \gamma_2/\gamma_3)/2 - k^2 k_+(1-\gamma_2 / \gamma_3)/2      - k_-   \beta^2 e^{-\imath 2 \theta} )$.    The spin-orbit strength $\alpha = -\imath 3 \gamma_3^2/2 m^2 [G_{LL}, k_z]$ is determined by the quantum well's confinement potential $V_c$.  It can be approximated analytically in a thin well with thickness $L$, where confinement creates a splitting $\Delta E \sim 2 \frac{\gamma_2}{m} \langle k_z^2 \rangle \propto 2 \frac{\gamma_2}{m} (2 \pi / L)^2$ between the heavy and light hole bands.   This energy scale justifies neglect of higher orders in the  potential $V_E$ and in $k_x, k_y$.   At leading order $\alpha = 6 \imath [V_E, k_z] (\gamma_3/ 2m \Delta E)^2$.  The $k_z$ appearing here is an operator and does not commute with the quantum well's built-in electric field; $[V_E,k_z] = - i e E$. Similar approximations determine that $\hat{k}_{HH}= \frac{k^2}{2 m_H}$, where the renormalized mass is $m_H=m/(\gamma_1 + \gamma_2 - 3 \gamma_3^2/\gamma_2)$. 

  The first term in $H_{\text{SO}}$ stands alone when there is neither strain nor anisotropy.  It is cubic in the spin-orbit strength and has f-wave character, reproducing the cubic dominance which is well known for  holes \cite{winkler2000}.  Optimal spin lifetimes are obtained only in the anisotropic limit $\gamma_2/\gamma_3 = -1$ where this term is entirely absent.  Anisotropy and strain produce the second and third terms, which respectively have Dresselhaus ($k_+$) and Rashba ($k_-$) character.  The spin-orbit interaction $H_{SO}$ has constant phase when the strain term's magnitude is tuned to match the magnitude of the anisotropy term, i.e. when  $\beta =  k_F \sqrt{(1- \gamma_2/\gamma_3)/2}$.  A truly constant phase is not achievable because $H_{SO}$ induces small anisotropies in the Fermi surface which are of order $E_{SO}/ E_F, \; E_{SO} = \alpha k_F^3$.   However when the strain is tuned properly these phase fluctuations are very small,  one component of the spin almost commutes with the Hamiltonian, and its lifetime becomes very large.

Strain can also tune  the Fermi surfaces of the spin $\uparrow$ and $\downarrow$ heavy holes to produce a quasi shifting symmetry (Eq. \ref{shifting}) that is key to persistent spin helices, which together with the quasi-conserved spin form a long-lived $SU(2)$ spin triplet.
 PSHs occur when the spin $\uparrow$ and $\downarrow$ Fermi surfaces, $\epsilon_{\uparrow,\downarrow}$,  have  identical shapes so that  a shift of $\vec{q} = \pm   \hat{Q}$ moves one Fermi surface on top of the other.   This shifting  symmetry can be written as:
\begin{equation}
\epsilon_{\downarrow}(\vec{k}) = \epsilon_{\uparrow}(\vec{k}
+\vec{Q}).\label{shifting}
\end{equation} \noindent  
Using our heavy hole Hamiltonian, Figure \ref{PSHHoleGasStrain}d shows that when the Fermi surfaces are tuned for degeneracy ($\beta = k_F$) they also obey the shifting symmetry that produces PSHs.  This is true both in the isotropic limit $\gamma_2 /  \gamma_3 = + 1$ and in the strongly anisotropic  limit $\gamma_2 /\gamma_3 = -1$.   In both limits the energy dispersion simplifies to    $E_\pm = (\vec{k} \pm \vec{Q}/2)^2/2m$ on the circle defined by $|\vec{k}| = k_F = \beta$.  Therefore  at leading order in the spin-orbit strength $E_{SO}/E_F$ the Fermi surfaces are circles offset from each other by $\pm \vec{Q}$, and produce  a spin helix standing wave.  The helix's   wave-vector  has magnitude $|\vec{Q}|  = 2 k_F E_{\text{SO}}/E_F$, is proportional to the spin-orbit strength, and is independent of scattering.

The  magnetoresistance is very sensitive to this physics.  When  the spin-orbit interaction has constant phase   the magnetoresistance  will become null or even change sign.  If the Fermi surfaces do not fulfill the nesting condition required by a   PSH then there will be neither weak localization nor antilocalization (null magnetoresistance).  If  a  PSH  exists then there will be  a complete reversal from weak localization to weak antilocalization, from negative to positive magnetoresistance.

 \begin{figure}[top]
\begin{center}
  \includegraphics[width=8cm, height=3cm, bb=0 0 480 220]{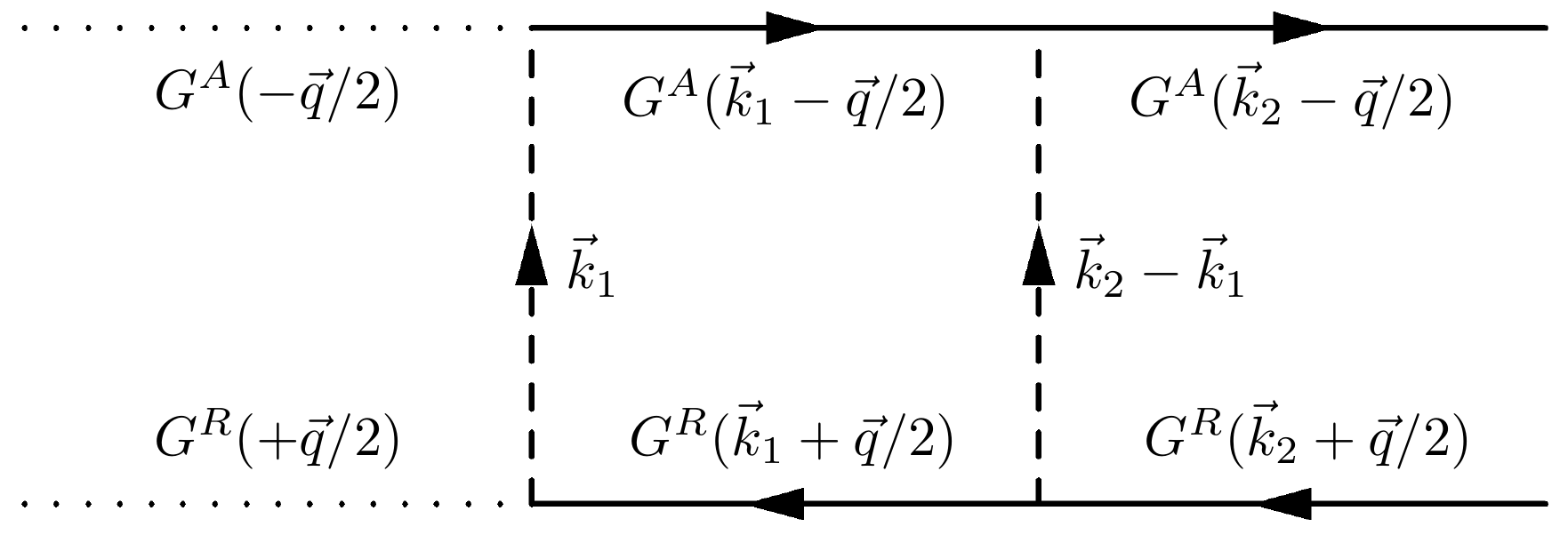} 
    \end{center}
    \caption{
Diagram illustrating the scattering that produces  spin-charge diffusion.  Here two scattering events are shown. $G^A$ and $G^R$  describe time evolution of the hole $\psi$ and its complex conjugate  $\psi^\dagger$.  Each scattering event causes correlations between $G^A$ and $G^R$ and is shown as a dashed line connecting the two. } 
    \label{fig:JointScattering}
\end{figure}
It may not be easy to observe these long spin lifetime effects, since semiconductors possess an approximate spherical symmetry \cite{Lipari70} which places  many of them near the isotropic limit $\gamma_2 / \gamma_3 = 1$.  However silicon is a notable exception, with $\gamma_2/\gamma_3 = 0.23$ \cite{LandoltBornstein17}.  Moreover  in many compounds there is considerable scatter in both experimental and theoretical estimates of $\gamma_2$ and $\gamma_3$, and certain authors have assigned GaP \cite{Vurgaftman01} , SiC \cite{Willatzen95}, and Boron-doped diamond \cite{Willatzen94} values of $\gamma_2/ \gamma_3 = 0.17, 0.24, $ and $-0.16$ respectively.   Lastly $\gamma_2 / \gamma_3$ remains completely unknown in the metallic phase of the pyrochlore iridates.   In these materials measurement of spin dynamics may prove to be a sensitive means of determining $\gamma_2/\gamma_3$.

 \begin{figure}[top]
\begin{center}
  \includegraphics[width=8.5cm,height=6.7cm]{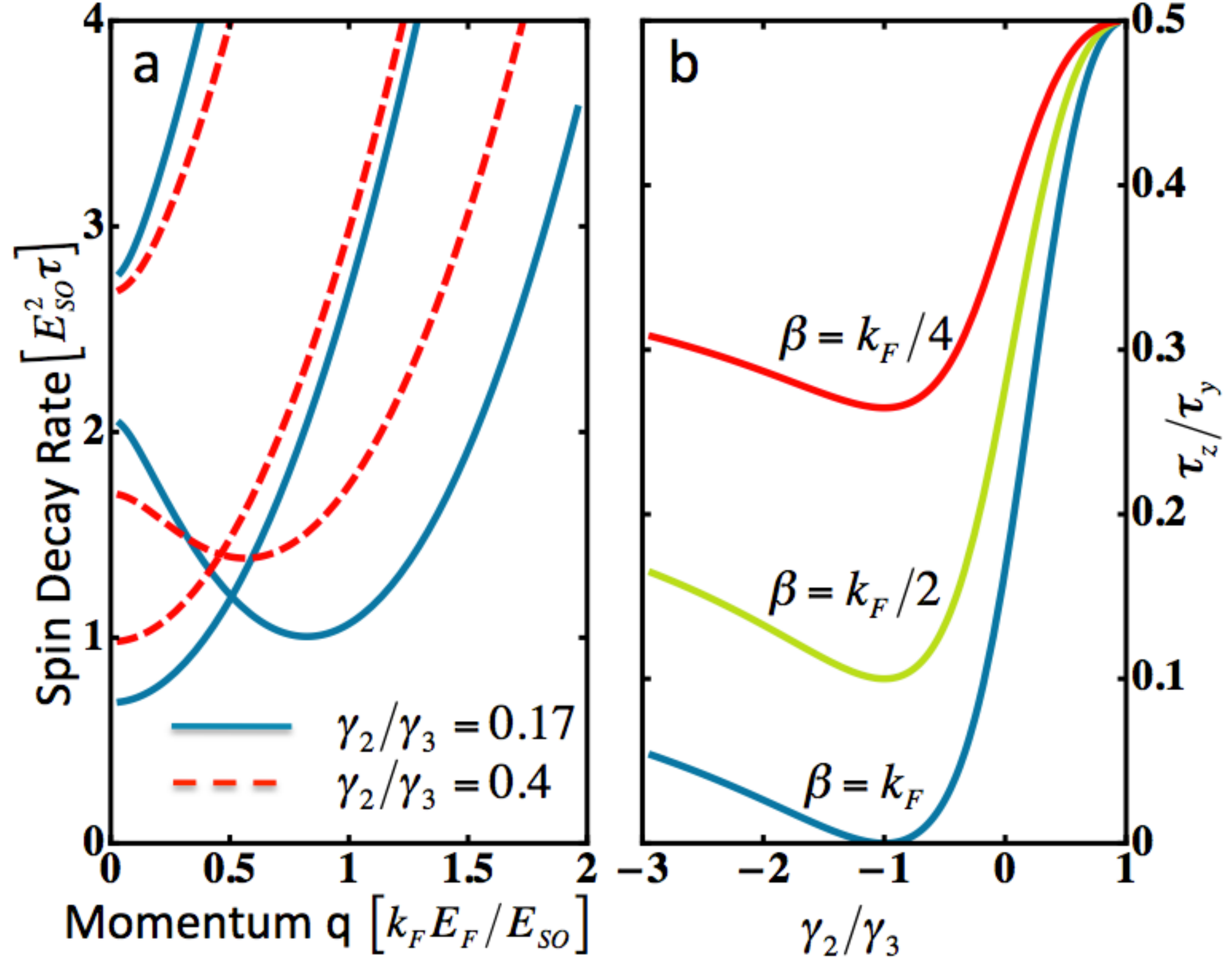} 
    \end{center}
    \caption{Strain-assisted suppression of spin decay.  Pane (a) shows the spin decay's dependence on momentum   when the strain is at its optimal value $\beta = k_F \sqrt{(1-\gamma_2/\gamma_3)/2}$.  Both $\vec{q}$ and  the strain are aligned with the $x$ axis.  Three decay rates are shown corresponding to $S_y$ and to linear combinations of $S_x$ and $S_z$.      $S_y$'s decay rate  is smallest at $q=0$,  while the $S_x, S_z$ rates are minimized at finite momenta $  \pm |\vec{Q}| = \pm (1- \gamma_2/\gamma_3) k_F E_F/ E_{\text{SO}}$.  These long-lived $\pm |\vec{Q}|$ excitations are the persistent spin helices.    The  minimum  decay rates of both $S_y$ and the PSHs go to zero when $\gamma_2/\gamma_3 = -1$, signaling that neither $S_y$ nor the PSHs decay.  Pane (b) illustrates this with the ratio $\tau_z/ \tau_y$ of  the decay times of spatially uniform  $S_z,  S_y$ spin distributions.  This ratio is proportional to the $S_y$ decay rate, and is zero at optimal $\gamma_2 / \gamma_3 = -1$ and optimal strain $\beta = k_F$.
       }
    \label{fig:LifetimeRatio}
\end{figure}

  Intrigued by this possibility, we study the equations of motion governing diffusion of the heavy holes, neglecting excitation and diffusion of light holes for analytic tractability.  The heavy holes form  a doublet with spin $\pm \frac{1}{2}$ and total angular momentum $j_z = \pm \frac{3}{2}$; we write the    charge density $N$ and the spin densities $S_i$ as a $4-$vector $\vec{\rho} = \begin{bmatrix} N, & S_x, & S_y, & S_z \end{bmatrix} $.   At time scales larger than the elastic scattering time $\tau$ their  diffusion and coupling to each other are controlled by the partial differential equation $\mathcal{D}^{-1} \vec{\rho} = 0$,   where the  $4 \times 4$ matrix  $\mathcal{D}_{ij}$ is called the diffuson.  
   We  derive  the diffuson using standard methods from the diagrammatic technique for disordered systems \cite{Hikami80,Burkov04,Wenk10}; details are reported in the supplementary material.   We model scattering with a  non-magnetic "white noise" disorder potential $V = \begin{bmatrix} 1 &  0 \\ 0 & 1 \end{bmatrix} u(\vec{r}), \, \langle u(\vec{r}) u(\acute{\vec{r}}) \rangle =  (2 \pi \nu \tau / \hbar)^{-1} \delta(\vec{r} - \acute{\vec{r}}) $, where $\nu$ is the density of states.   We assume as usual that the Fermi surface is dominant ($E_F \tau / \hbar \gg 1$). The diffuson describes sequences of events in which the hole wave-function $\psi$ and its conjugate $\psi^\dagger$ move together, scattering in unison.   Two scattering events are  pictured in Figure ~\ref{fig:JointScattering}.  A single  scattering  is described by the operator $I_{ij}$, and the diffuson  sums diagrams with any number of scatterings;  $ \mathcal{D}(\vec{q}, \omega)  =  \sum_{n=0}^\infty (I_{ij})^n = (1 - I_{ij})^{-1}$. $I_{ij}$ is given by the integral 
\begin{equation} 
\begin{split}
I_{ij} = \frac{\hbar}{4 \pi \nu \tau} \int d\vec{k} \, {Tr}(\, G^A( \vec{k} - \vec{q}/2, E_F)  \\ \sigma_i \, G^R(  \vec{k} + \vec{q}/2,  E_F + \hbar \omega) \, \sigma_j \, ) 
\end{split}
\label{JointScatteringOperator}
\end{equation}
$G^A$ and $G^R$  are the disorder-averaged single-particle Green's function and $\vec{q}$ is the diffuson momentum.  The trace is taken over the spin indices of $G^A, G^R, \sigma_i,$ and $\sigma_j$, which are all $2 \times 2$ matrices in spin space.

 We here report the diffusion equations at leading order in the spin orbit strength and in the momentum, with strain along the $x$ axis:
\begin{eqnarray}
\partial_t N & = & D \nabla^2 N 
\nonumber \\ 
\partial_t S_x &=& D \nabla^2 S_x - (C_1 +  C_2) \partial_x S_z  - (\frac{1}{T} + U)S_x 
\nonumber \\ 
\partial_t S_y & = & D \nabla^2 S_y - (C_1 -  C_2)\partial_y S_z- (\frac{1}{T}- U)S_y 
\\ \nonumber  
  \partial_t S_z & =  &D \nabla^2 S_z + (C_1 +  C_2) \partial_x S_x + (C_1 -  C_2)\partial_y S_y-\frac{2}{T} S_z
\end{eqnarray}
\noindent where the coefficients read:
\begin{eqnarray} \label{Coeffs}
&U = (1 - (\gamma_2/\gamma_3)) \, 2 \alpha^2 k_F^4  \beta^2  \tau;   \
\\ \nonumber 
&\frac{1}{T} =  \alpha^2 k_F^2 (k_F^4(1 + \gamma_2^2/\gamma_3^2) + 2\beta^4)\tau;   \
\\ \nonumber 
 &C_1 =4 \alpha \beta^2   E_F  \tau, \;\;\;   C_2 = (1 - (\gamma_2/\gamma_3))\, 2 \alpha k_F^2   E_F  \tau
 \label{DiffusionEq}
\end{eqnarray}
\noindent  $D = v_F^2 \tau/2$ is the usual diffusion constant. The spin-spin couplings $C_1$ and $C_2$ are caused by respectively strain and anisotropy, while the lifetime splitting $U$ is caused by both anisotropy and strain together.   We have checked that higher order terms do not cause qualitative changes in the spin lifetime or the spin-spin couplings, although they do produce a small spin-charge coupling.    When the strain is dominant  ($\beta/k_F \rightarrow \infty, \alpha \cdot \beta^2 \propto 1$) we obtain the well known  Rashba spin diffusion equations ~\cite{Burkov04, Mishchenko2004a}.   The couplings $C_1 \propto \gamma_3, \, C_2 \propto \gamma_3 (\gamma_3 - \gamma_2)$, lifetime $1/T \propto \gamma_3^2(\gamma_2^2 + \gamma_3^2)$, and lifetime splitting   $U \propto \gamma_3^2 (\gamma_3 - \gamma_2)$ are all highly sensitive to the Luttinger parameters, whose numerical values remain controversial.  Experimental measurements of the spin dynamics should help to determine the Luttinger parameters.
 
Lastly we discuss the hole spin helix, a spin density wave  aligned with  the $x$ axis, precessing in the $x-z$ plane. If  $\gamma_2/ \gamma_3 < 0.6$ then adjusting the strain strength  to $\beta =  k_F \sqrt{(1- \gamma_2/\gamma_3)/2}$ produces an optimal spin helix lifetime $T_{PSH}^{-1} = 3/2 \, \, (1+\gamma_2/\gamma_3)^2 E_{\text{SO}}^2 \tau / 2$.    The solid blue lines in figure \ref{fig:LifetimeRatio}a   illustrate the decay rates in GaP:    there are two spin helices with enhanced lifetimes at opposite wave-vectors $\pm \vec{Q}, \; |\vec{Q}| \sim(1 - \gamma_2 / \gamma_3)   k_F  E_{\text{SO}}/ E_F$.  Accompanying the spin helices, the $S_y$ spin component also exhibits an enhanced lifetime  $T_y = (3/2) T_{PSH}$.   As discussed earlier, Fermi surface anisotropy caps this lifetime at order $O(E_F^2 / E_{SO}^4 \tau)$. The longest lifetime coincides with  $C_1 = C_2, \; U=1/T, \;C^2=8DU$.     Figure \ref{fig:LifetimeRatio}b shows the contrast ratio of the $S_z$ lifetime to the $S_y$ lifetime, which is $1/2$ in the isotropic limit.  This ratio is reduced by a factor of two to $\sim 1/4$ when $\gamma_2/\gamma_3 \sim 0.2$ (Si, GaP, and SiC).  The corresponding hole spin helix  lifetime enhancement is $ \sim 8/3$.  If the prediction $\gamma_2/\gamma_3 = -0.16$ for Boron-doped diamond \cite{Willatzen94} is correct then the  hole spin helix's non-uniform lifetime enhancement would reach $6.4$.  This can be confirmed  by transient spin grating spectroscopy.

\acknowledgments{We acknowledge useful discussions with R. Winkler, M. Shayegan, B. Normand, A. MacDonald, and D. Culcer.  B.A.B. acknowledges the hospitality of the Yukawa Institute for Theoretical Physics, specifically the Spin Transport in Condensed Matter workshop in November 2008, during which the current work was begun. B.A.B. was supported by ONR- N00014-11-1-0635, Darpa-N66001-11-1-4110, David and Lucile Packard Foundation, and MURI-130-6082. This work was supported by the National Science Foundation of China and by the 973 program of China  under Contract No. 2011CBA00108. V.E.S. thanks Xi Dai and the IOP, which hosted and supported almost all of his work.}

\begin{appendix}
\begin{widetext}
\section{Approximate Form of the Heavy Hole Effective Hamiltonian}
The four-band Luttinger model \cite{Luttinger56, Winkler03, Lu05}, with the Bir-Pikus strain Hamiltonian, in a quantum well extended along the $x - y$ plane and grown along the $001$ ($z$) axis, is represented in the $J_z = +3/2, -3/2, +1/2, -1/2$ basis as:
\begin{eqnarray}
V(z) & + &    \begin{bmatrix} k_{HH} & 0 & S & R^*  \\ 0 & k_{HH}  & R & -S^*  \\ S^* & R^* &  k_{LL}  & 0  \\ R  & -S & 0 & k_{LL}  \end{bmatrix}
\nonumber \\
k_{HH} &=& (\gamma_1 + \gamma_2)k^2/2m + (\gamma_1 - 2 \gamma_2)k_z^2/2m  -a (\epsilon_{xx} + \epsilon_{yy} + \epsilon_{zz}) +b(\epsilon_{zz} - \epsilon_{xx}/2 - \epsilon_{yy}/2)
\nonumber \\
k_{LL} & = & (\gamma_1 - \gamma_2)k^2/2m + (\gamma_1 + 2 \gamma_2)k_z^2/2m   -a (\epsilon_{xx} + \epsilon_{yy} + \epsilon_{zz}) -b(\epsilon_{zz} - \epsilon_{xx}/2 - \epsilon_{yy}/2)
\nonumber \\
S & = & -  \sqrt{3} \gamma_3 k_- k_z/ m  -  d ( \imath \epsilon_{zy} -  \epsilon_{zx})
 \nonumber \\
R & = &   \sqrt{3} \gamma_2 (k_y^2 - k_x^2) / 2m - \imath \sqrt{3} \gamma_3 k_x k_y / m  + \sqrt{3}b (\epsilon_{xx} - \epsilon_{yy}) /2+ \imath d \epsilon_{xy}
 \nonumber \\
 & = &   -\frac{\sqrt{3}}{4m} (k_+^2 (\gamma_2 + \gamma_3) + k_-^2 (\gamma_2 - \gamma_3) -  2 \gamma_3 \beta^2 e^{\imath 2 \theta}), \;  \beta^2 e^{\imath 2 \theta} = \frac{ m b}{ \gamma_3} (\epsilon_{xx} - \epsilon_{yy})   + \imath  \frac{2 m d} {\sqrt{3} \gamma_3 }\epsilon_{xy}  
 \nonumber \\
 k_+ &=& k_x + \imath k_y, \; k_- = k_x - \imath k_y
\end{eqnarray}
$m$ is the electron mass, $\gamma_1, \gamma_2, \gamma_3$ are the material-specific Luttinger Hamiltonian parameters, $a,b,d$ are the material-specific strain deformation potentials, and the $\epsilon$ parameters describe the strain on the sample.  $V(z) = V_c + V_E $ is the sum of the quantum well's confinement potential $V_c(z)$ (which is symmetric under inversions) and a symmetry-breaking term $ V_E  = - e E z$.   $k_{HH}$ and $k_{LL}$  are kinetic operators for the heavy and light holes respectively.  $S$ couples  states whose spin have the same sign, while $R$ couples states whose spin have opposite sign.

Hooke's law implies that in-plane strain $\epsilon_{xx}, \epsilon_{yy}, \epsilon_{xy}$ applied to the quantum well will cause out-of-plane strain as well. \footnote{Many thanks to Roland Winkler for explaining these points about Hooke's law.}  When the crystal growth is along the 001 direction this physics simplifies and produces only one extra strain component, $\epsilon_{zz}$.  The only effect of $\epsilon_{zz}$ is a shift in the splitting between the light and heavy holes.  For other growth directions $\epsilon_{zx}, \epsilon_{zy}$ contribute to the $S$ matrix element, but we will show that these terms make no contribution to the commutator which controls the spin orbit interaction.

Since we are only interested in observables constructed from heavy holes, we can apply a unitary transformation to the light holes: $ | +1/2 \rangle  \rightarrow (k_+ / k )  | +1/2 \rangle, \;    |  -1/2 \rangle  \rightarrow (k_- / k )  |  -1/2 \rangle$.  As a result, the Hamiltonian transforms:
$ S \rightarrow  (k_+ / k )  S=- \sqrt{3} \gamma_3 k k_z/ m  -   (k_+ / k )d (\imath \epsilon_{zy} - \epsilon_{zx}), \  R \rightarrow  (k_+ / k ) R =  -\frac{\sqrt{3}}{4m} ((k_+^3 / k) (\gamma_2 + \gamma_3) + k_- k (\gamma_2 - \gamma_3) -( k_+ /k) 2 \gamma_3  \beta^2 e^{\imath 2 \theta})$.

The second and third terms in $R$ combine to have constant phase when $\beta^2 =  k_F^2 (1 - \gamma_2 / \gamma_3) $, and the first term in $R$ can be completely eliminated by setting $\gamma_2 / \gamma_3 = -1$, rendering $R$ completely real.  In momentum space the only remaining complex term in the Luttinger Hamiltonian is  $S$'s strain term $- \imath  (k_+ / k ) d  \epsilon_{zy}$.  

We will now prove that  $S$'s strain term $- \imath  (k_+ / k ) d ( \epsilon_{zy}+ \imath \epsilon_{zx}) $ has no effect on the coupling between the heavy holes.   The effective Hamiltonian for the heavy holes is:
\begin{eqnarray}
H_{HH}&=& V(z) +  \begin{bmatrix} k_{HH}  & 0 \\ 0 & K_{HH} \end{bmatrix} - \begin{bmatrix} S  & R^* \\ R & -S^* \end{bmatrix} (k_{LL} + V(z) - E)^{-1} \begin{bmatrix} S^*  & R^* \\ R & -S \end{bmatrix}
\nonumber \\
& = &  V(z)  +   \begin{bmatrix} \acute{k}_{HH}  &   [ (k_{LL} + V(z) - E)^{-1}, S] R^* 
 \\   - [(k_{LL} + V(z) - E)^{-1} , S^*]  R & \acute{K}_{HH}
\end{bmatrix}
\nonumber \\
\acute{k}_{HH} &= & k_{HH} - S (k_{LL} + V(z) - E)^{-1} S^* - R^* (k_{LL} + V(z) - E)^{-1} R
\end{eqnarray}
The strain term $- \imath  (k_+ / k ) d ( \epsilon_{zy}+ \imath \epsilon_{zx})$ has no $z$ dependence, so it commutes with $k_{LL}$ and $V(z)$, and makes no contribution to the commutator.   In fact the commutator reduces to  $- \sqrt{3} \gamma_3 k / m [ (k_{LL} + V(z) - E)^{-1}, k_z]$.  

After transforming  to position space the commutator is $-\imath \sqrt{3} \gamma_3 k / m [ (k_{LL} + V(z) - E)^{-1}, \partial_z]$.  Its phase is manifestly constant.   This proves that  the off-diagonal  elements of the heavy hole Hamiltonian $H_{HH}$  have (up to a factor of $\imath$) the same phase as $R$.    If $R$'s phase is constant then the spin-orbit interaction also has constant phase and one component of the  spin is conserved.  This is an exact result for the full four-band Luttinger Hamiltonian.

Moreover we note that diagonal elements of $H_{HH}$ are proportional to the identity; any splitting between the heavy holes comes only from the off-diagonal elements.

The previous steps were exact.  Now we obtain a simple but approximate form for the heavy-hole Hamiltonian.  We  simplify the effective Hamiltonian $H_{HH}$ by assuming that the quantum well is thin, and that the confinement potential $V_c(z)$ along the $z$ axis  splits the spectrum into 2-D sub bands corresponding to eigenstates of $V_c$.   We assume that the Fermi level is used to regulate the  hole carrier density to a small value where only the first 2-D sub-band contributes to transport.  Therefore we can assume that the system is in the lowest eigenstate of $V_c(z)$, which can be be replaced everywhere by its lowest eigenvalue.  

 Since the quantum well is thin, $ k_z \gg k_x, k_y$ and therefore $R$'s contribution to the diagonal is negligible.  Next we assume that the charge asymmetry is small compared to the splitting between light and heavy holes  ($V_E \ll k_{LL} - E$), we expand in powers of $V_E$, and we evaluate the expectation value $\langle k_{LL} - E \rangle =  \Delta E$. 
\begin{eqnarray}
H_{HH}&=&  V(z)  +   \begin{bmatrix} \acute{k}_{HH} - S S^* (\Delta E)^{-1}   & -\imath \alpha  \frac{2m}{\sqrt{3}  \gamma_3}  k  R^*
 \\    \imath  \alpha \frac{2m}{\sqrt{3}  \gamma_3}  k  R & \acute{k}_{HH} - S S^* (\Delta E)^{-1} 
\end{bmatrix}
\nonumber  \\
 \alpha  &=&  \imath 6 \gamma_3^2 [V_E, k_z]/(2 m \Delta E)^2, \; [V_E, k_z ] = -\imath E_z 
\end{eqnarray}
$\alpha$ is the strength of the heavy hole spin-orbit interaction.  Lastly we approximate the diagonal elements of the effective Hamiltonian by  ignoring terms which are independent of $k_+, k_-$ and finding the effective mass $m_H$ which controls the diagonal's in-plane kinetic energy $k^2 / 2 m_H$.  This requires calculation of the expectation value of $ \langle k_z^2 \rangle$, which we perform using  $S S^* \approx 3 \gamma_3^2  k^2 \langle k_z^2 \rangle / m^2$.   We  use $ \Delta E = \langle k_{LH} - k_{HH} \rangle \approx \frac{ 2 \gamma_2}{m} \langle k_z^2 \rangle  $ and obtain the final effective Hamiltonian of the heavy holes:
\begin{eqnarray}
H_{HH}&=&   \begin{bmatrix} k^2 / 2 m_H  & -\imath \alpha   \frac{2m}{\sqrt{3}  \gamma_3}  k  R^*
 \\    \imath \alpha   \frac{2m}{\sqrt{3}  \gamma_3}  k  R & k^2 / 2 m_H
\end{bmatrix}
\\ \nonumber 
&=&  \begin{bmatrix} k^2 / 2 m_H  & \imath \alpha  (k_-^3 (1+\gamma_2/  \gamma_3 )/2 - k^2 k_+ (1-\gamma_2 / \gamma_3 )/2 -  k_- \beta^2 e^{-\imath 2 \theta})
 \\  - \imath \alpha  (k_+^3 (1+\gamma_2/  \gamma_3 )/2 - k^2 k_- (1-\gamma_2 / \gamma_3 )/2 -  k_+ \beta^2 e^{\imath 2 \theta})& k^2 / 2 m_H
\end{bmatrix}
\nonumber \\
 &=& k^2/2 m_H + a_x \sigma_x + a_y \sigma_y, \; 
 \nonumber \\
  a_x   &=& - \alpha  k_y(- 2 k_x^2  (1+\gamma_2/  \gamma_3 ) + k^2   (\gamma_2/  \gamma_3 )) -  \alpha  \beta^2  (k_x \sin 2 \theta + k_y \cos 2 \theta)
 \nonumber \\
            a_y    &=&  -  \alpha  k_x(- 2  k_y^2 (1+\gamma_2/  \gamma_3 ) + k^2  (\gamma_2/  \gamma_3 ) )+  \alpha  \beta^2  (k_x \cos 2 \theta - k_y \sin 2 \theta)
\end{eqnarray}
The  renormalized in-plane mass is $ m_H= m/(\gamma_1 + \gamma_2 - 3 \gamma_3^2/\gamma_2)$.

\section{Derivation of The Spin Diffusion Equations}

We study the equations of motion governing diffusion of the heavy holes, neglecting excitation and diffusion of light holes.  The heavy holes form  a doublet with spin $\pm \frac{1}{2}$ and total angular momentum $j_z = \pm \frac{3}{2}$; we write the    charge density $N$ and the spin densities $S_i$ as a $4-$vector $\vec{\rho} = \begin{bmatrix} N, & S_x, & S_y, & S_z \end{bmatrix} $.     At time scales larger than the elastic scattering time $\tau$ their  diffusion and coupling to each other are controlled by the partial differential equation $\mathcal{D}_{ij}^{-1} \vec{\rho} = 0$,   where the  $4 \times 4$ matrix  $\mathcal{D}_{ij}$ is called the diffuson and is determined by  $ \mathcal{D}_{ij}(\vec{q}, \omega)  = (1 - I)^{-1}$. The joint scattering operator $I_{ij}$ is  is given by the integral 
\begin{equation} 
I_{ij} = \frac{\hbar}{4 \pi \nu \tau} \int d\vec{k} \, {Tr}(\, G^A( \vec{k} - \vec{q}/2, E_F)  \\ \sigma_i \, G^R(  \vec{k} + \vec{q}/2,  E_F + \hbar \omega) \, \sigma_j \, ) 
\label{JointScatteringOperator}
\end{equation}
$G^A$ and $G^R$  are the disorder-averaged single-particle Green's functions which express uncorrelated movements of $\psi$ and $\psi^\dagger$, while $\vec{q}$ is the diffuson momentum.  The trace is taken over the spin indices of $G^A, G^R, \sigma_i,$ and $\sigma_j$, which are all $2 \times 2$ matrices in spin space.  

We have computed the diffuson  systematically to next to leading order in $1/E_F \tau$ and to fourth order in $E_{SO}/E_F$.  This involved going to fourth order in $E_F \tau, q,$ and $ E_{SO}/E_F$.    However the dominant physics is already visible at second order.  At this order the diffuson simplifies to:
\begin{eqnarray}
 \mathcal{D}^{-1}  &  = & 1- \frac{1}{ 16 \pi  (1 - \imath \omega \tau)}\sum_{s,\acute{s},\nu= \pm 1}   \int d\theta  M \;  \frac{1}{1 - \imath \tau \nu \hbar^{-1} (E_F -  E(\vec{k} +  \nu \vec{q} , (1 - \nu)s/2 + (1 + \nu)\acute{s}/2) )}, 
 \nonumber \\
E_F - E & \approx & -2 \nu q \cos(\theta - \theta_q)  E_F/k_F 
+  \nu (s- \acute{s})\sqrt{a_x^2 + a_y^2}
   \nonumber \\
 M & = &  \begin{bmatrix}1 + s \acute{s}  & (s  + \acute{s}) \cos \phi &   (s + \acute{s} ) \sin \phi & 0  \\ (s  + \acute{s} ) \cos \phi & 1 + s \acute{s} \cos(2 \phi)  & s \acute{s} \sin(2 \phi) & -\imath (s - \acute{s}) \sin \phi \\  (s  + \acute{s} ) \sin \phi & s \acute{s} \sin(2 \phi)  &1-s \acute{s} \cos(2 \phi) & \imath (s - \acute{s}) \cos \phi  \\ 0 & \imath (s - \acute{s}) \sin \phi & - \imath ( s-  \acute{s}) \cos \phi & 1 - s \acute{s}  \end{bmatrix}
\nonumber \\
\cos \phi & = & a_x(\vec{k})/ \sqrt{a_x^2 + a_y^2}, \; \sin \phi  =  \pm \sqrt{1 - \cos^2 \phi} = a_y(\vec{k})/\sqrt{a_x^2 + a_y^2},  
\nonumber \\
\vec{k} & = & k_F(\cos \theta \hat{x} +  \sin \theta \hat{y}), \vec{q}  = q(\cos \theta_q \hat{x} +  \sin \theta_q \hat{y})
\end{eqnarray}

After some algebra we obtain: 
\begin{eqnarray}
 \mathcal{D}^{-1} \tau^{-1}  &  = &  
-\imath \omega  + D q^2  + \begin{bmatrix} 0 & 0 & 0 & 0 \\ 0 & U  \cos 2\theta+ 1/T & U  \sin 2\theta & - \imath (C_1 q \cos(\theta_q + 2 \theta) + C_2 q_x)  \\ 0 & U  \sin 2\theta &  - U \cos 2 \theta + 1/ T &  -\imath (C_1 q \sin(\theta_q  + 2 \theta) - C_2   q_y)    \\ 0 &   \imath (C_1 q \cos(\theta_q + 2 \theta) + C_2 q_x)      &  \imath (C_1 q \sin(\theta_q  + 2 \theta) - C_2   q_y) &  2/ T \end{bmatrix}
\nonumber \\
1/T & = &   k_F^2 (\alpha  / \hbar)^2   \tau (k_F^4(1 + (\gamma_2/\gamma_3)^2) + 2 \beta^4)   \propto (E_{SO} \tau / \hbar)^2 / \tau
\nonumber \\
D & = & 2 (E_F \tau / \hbar k_F)^2 / \tau = v_F^2  \tau / 2 \propto (E_{SO} \tau / \hbar)^2 (E_F / E_{SO} k_F)^2  / \tau
\nonumber \\
U & = &  2 (\alpha  / \hbar)^2  \beta^2 k_F^4  \tau (1 - (\gamma_2/\gamma_3))   \propto (E_{SO} \tau / \hbar)^2 / \tau
\nonumber \\
 C_1 &=& 4 \alpha \beta^2   E_F  \tau /\hbar^2 \propto (E_{SO} \tau / \hbar)^2 (E_F / E_{SO} k_F)  / \tau, \;\;\;   
 \nonumber \\
 C_2 &=& (1 - (\gamma_2/\gamma_3))\, 2 \alpha k_F^2   E_F \tau /\hbar^2 \propto (E_{SO} \tau / \hbar)^2 (E_F / E_{SO} k_F) / \tau
\end{eqnarray}

Going to higher order, we found that all the non-zero elements of $ \mathcal{D}^{-1}$ have corrections.  There is also a spin-charge coupling at higher order, but it is small even compared to the corrections which we just mentioned.

 Assuming the time-dependence $\exp{(-i \omega t -  i \vec{q}\cdot \vec{r})}$, we obtain the equations of motion:
\begin{eqnarray}
\partial_t N & = & D \nabla^2 N 
\nonumber \\ 
\partial_t S_x &=& D \nabla^2 S_x     - (U  \cos 2\theta + \frac{1}{T})S_x - U  \sin 2\theta \, S_y  -  C_2 \partial_x S_z - C_1 (\cos 2 \theta  \, \partial_x - \sin 2 \theta  \,  \partial_y )S_z
\nonumber \\ 
\partial_t S_y & = & D \nabla^2 S_y - U  \sin 2\theta \, S_x - (-U  \cos 2\theta +\frac{1}{T})S_y    +  C_2 \partial_y S_z - C_1 (\sin 2 \theta \,  \partial_x + \cos 2 \theta  \, \partial_y )S_z
\nonumber \\ 
  \partial_t S_z & =  &D \nabla^2 S_z -\frac{2}{T} S_z     +  C_2 (\partial_x S_x  - \partial_y S_y) +  C_1 (\cos 2 \theta  \, \partial_x - \sin 2 \theta  \, \partial_y )S_x + C_1 (\sin 2 \theta  \,  \partial_x + \cos 2 \theta  \,  \partial_y )S_y
\end{eqnarray}
 
When the strain is along the $x$ axis ($\theta = 0$) this simplifies to:
\begin{eqnarray}
\partial_t N & = & D \nabla^2 N 
\nonumber \\ 
\partial_t S_x &=& D \nabla^2 S_x     - (U   + \frac{1}{T})S_x   -  C_2 \partial_x S_z - C_1  \partial_x S_z
\nonumber \\ 
\partial_t S_y & = & D \nabla^2 S_y  - (-U   +\frac{1}{T})S_y    +  C_2 \partial_y S_z - C_1  \partial_y S_z
\nonumber \\ 
  \partial_t S_z & =  &D \nabla^2 S_z -\frac{2}{T} S_z     +  C_2 (\partial_x S_x  - \partial_y S_y) +  C_1  \partial_x S_x + C_1   \partial_y S_y
\end{eqnarray}
\end{widetext}
\end{appendix}

\bibliography{PSHHole}
\end{document}